\documentclass[fleqn,twoside]{article}
\usepackage{espcrc2}

\usepackage{graphicx}
\newcommand{\AmS}{{\protect\the\textfont2
  A\kern-.1667em\lower.5ex\hbox{M}\kern-.125emS}}

\title{Radiative corrections to the Casimir effect
for the massive scalar field}

\author{F. A. Barone\address[UFRJ]{Instituto de F\'{\i}sica,
Universidade Federal do Rio de Janeiro, \\
Caixa Postal 68528, 21941-972 Rio de Janeiro, RJ, Brazil}\thanks{E-mail: fabricio@if.ufrj.br},
        R. M. Cavalcanti\addressmark\thanks{E-mail: rmoritz@if.ufrj.br}, and
        C. Farina\addressmark\thanks{E-mail: farina@if.ufrj.br}.}
      
\begin{document}

\begin{abstract}
We compute the first radiative correction to the Casimir energy of a 
massive scalar field with a $\lambda\phi^4$ self-interaction in the presence 
of two parallel plates. Three kinds of boundary conditions are considered: 
Dirichlet-Dirichlet, Neumann-Neumann and Dirichlet-Neumann. 
We use dimensional and analytical regularizations to obtain our physical results. 
\vspace{1pc}
\end{abstract}

\maketitle

\section{INTRODUCTION}

In 1948 Casimir predicted an unexpected quantum field
theory (QFT) effect: as a consequence of distorting the
vacuum of the quantized electromagnetic field,
two conducting and neutral parallel plates should
attract each other with a force proportional to the
inverse of the fourth power of their separation \cite{Casimir1948}
(for general reviews on the Casimir effect, see 
Refs.\ \cite{Revisoes}; for an introductory guide, see Ref.\ \cite{Lamoreaux}).
The first attempt to observe this phenomenon was made only in 1958 by 
Sparnaay \cite{Sparnaay1958}. However, the accuracy 
achieved by Sparnaay allowed him only to conclude that the experimental data 
was compatible with Casimir's theoretical predictions. For different reasons, 
new experiments involving directly metal bodies were performed only very 
recently \cite{experiments}, but this time 
the high accuracy obtained  and the excellent agreement between the experimental 
data and theory permit us to state safely that the Casimir effect is well 
established nowadays.

The Casimir effect is not a peculiarity of the quantized electromagnetic 
field. In fact, the vacuum state (and its energy) of any relativistic quantum field, bosonic or 
fermionic, depends on the boundary conditions (BC) imposed on the fields or,
more generally, on the classical background with which the fields interact.
This makes the Casimir effect an important topic of research, with
applications in many branches of physics
(see Bordag {\it et al.}\ in \cite{Revisoes}).

Even though QFT is mainly concerned with interacting fields, most papers on the 
Casimir effect deals with non-interacting fields. The computation of radiative
corrections to the Casimir energy of interacting fields has been performed
in relatively few papers. (See, for instance, Refs.\ \cite{QED} in the context
of QED, and Refs.\ \cite{Scalar,LR,FRF} for scalar fields.)

For non-interacting fields, the Casimir energy is given by the sum,
properly regularized and renormalized, of the zero-point energy
of the normal modes of the fields, which behave as a collection
of independent harmonic oscillators. Hence, at the one-loop 
level the Casimir energy is sensitive only to the fields' eigenfrequencies, 
but not to their eigenmodes. This is the reason why, for instance, the Casimir
energy of a scalar field confined between two parallel plates is the
same for Dirichlet or Neumann BC on both plates.\footnote{To be precise,
there are modes in the Neumann BC case that are not present in the
Dirichlet BC case, but since their zero-point energy does not depend
on the distance between the plates, they do not contribute to the Casimir
force and can be discarded.}
In the case of interacting fields the situation is more complicated:
the independent harmonic oscillators become anharmonic and coupled when
the interaction is turned on. Therefore, one has to take into
account not only the oscillators' zero-point energy --- which are
modified by the interaction, as in the Lamb shift ---, but also
the interaction energy among the oscillators \cite{LR}. This should
made the radiative corrections to the Casimir energy sensitive
to the form of the eigenmodes; in particular, it should depend
on the BC imposed on the fields. In spite of this, Krech and 
Dietrich \cite{Scalar} showed that the $O(\lambda)$ correction 
to the Casimir energy of the massless $\lambda\phi^4$ theory   
is the same for Dirichlet and Neumann BC on a pair of plates.

Our purpose here is to report the results of our investigation \cite{FRF}
on whether that equality is also true for a massive field.
We computed the $O(\lambda)$
radiative correction to the Casimir energy of the massive $\lambda\phi^4$ theory 
subject to Dirichlet and Neumann BC on a pair of parallel plates.
Our results show that the mentioned equality is valid only in
the massless cases. We also extend our calculations to the case in which
the field is subject to Dirichlet BC on one of the plates and to Neumann BC on
the other one.

\section{ONE-LOOP CASIMIR EFFECT}

In order to introduce some notation and basic ideas, we briefly sketch 
in this section some results for the Casimir energy of a non-interacting 
massive scalar field submitted to 
BC at $z=0$ and $z=a$.\footnote{Conventions: $\hbar=c=1$, $x=(\tau,{\bf r},z)$.}
We shall consider three distinct BC, denoted  by DD, NN and DN, and given, 
respectively, by: (i) $\phi(z=0)=\phi(z=a)=0$; (ii) 
$\partial\phi/\partial z\vert_{z=0}
=\partial\phi/\partial z\vert_{z=a}=0$, and (iii)
$\phi(z=0)=\partial\phi/\partial z\vert_{z=a}=0$.

The Casimir energy per unit area when Dirichlet or Neumann  
BC are used on the two planes is given by 
\begin{equation}
\label{E0}
E_{\rm DD}^{(0)}=E_{\rm NN}^{(0)}=-\frac{m^2}{8\pi^2 a}
\sum_{n=1}^{\infty}\frac{K_2(2nma)}{n^2}.
\end{equation}
For the mixed BC (Dirichlet-Neumann), we get
\begin{eqnarray}
\label{E0DN}
E_{\rm DN}^{(0)}\!\!\!\!&=&\!\!\!\!-\frac{m^2}{16\pi^2a}
\nonumber \\
\!\!\!\!& &\!\!\!\!\times\sum_{n=1}^{\infty} \frac{1}{n^2}\,[K_2(4nma)- 2K_2(2nma)].
\end{eqnarray}
The small mass limit ($ma\ll 1$) of these expressions is given by
\begin{equation}
\label{DDmassanula}
E_{\rm DD}^{(0)}=E_{\rm NN}^{(0)}=
-\frac{1}{1440}\frac{\pi^2}{a^3} + \frac{1}{96}\frac{m^2}{a}+O(m^3),
\end{equation}
\begin{equation}
E_{\rm DN}^{(0)}={7\over 8}\frac{\pi^2}{1440}{1\over a^3}-
{1\over 192}{m^2\over a}+O(m^3).
\label{DNmassanula}
\end{equation}
On the other hand, in the large mass limit ($ma\gg 1$) Eqs.\ (\ref{E0})
and (\ref{E0DN}) yield
\begin{equation}
\label{EDD0m}
E_{\rm DD}^{(0)}=E_{\rm NN}^{(0)}\approx - \frac{1}{16} 
\left(\frac{m}{\pi a}\right)^{3/2}\,\exp(-2ma),
\end{equation}
\begin{equation}
\label{EDN0m}
E_{\rm DN}^{(0)}\approx \frac{1}{16}\left(\frac{m}{\pi a} \right)^{3/2} 
\,\exp(-2ma).
\end{equation}
It is worth noting that the first term on the r.h.s.\ of Eq.\ (\ref{DDmassanula}) is 
precisely half the Casimir energy per unit area for the electromagnetic field 
between two perfectly conducting (or infinitely permeable)
parallel plates \cite{Casimir1948}, while the first term 
on the r.h.s.\ of Eq.\ (\ref{DNmassanula}) is half   the Casimir energy per unit 
area for the electromagnetic field between a perfectly conducting plate and 
an infinitely permeable one \cite{Boyer74Hushwater97CougoPinto99}.

\section{TWO-LOOP CASIMIR EFFECT}

Now we shall consider the $\lambda\phi^4$ model, defined by the Euclidean 
Lagrangian density
\begin {equation}
{\cal L}_{\rm E}={1\over 2}\,(\partial_{\mu}\phi)^{2}+{1\over 2}\,m^{2}\phi^{2}
+{\lambda\over 4!}\,\phi^{4}+{\cal L}_{\rm CT},
\end {equation}
where ${\cal L}_{\rm CT}$ contains the usual  renormalization counterterms. The 
interacting field is submitted to one of the three BC already considered before.
(However, we shall present the calculations only for the DN case,
as the other two BC were considered in detail in \cite{FRF}.)
Using perturbation theory,  the $O(\lambda)$ correction 
to the previous results can be written as
\begin {eqnarray}
\label {Einicial}
E^{(1)}\!\!\!\!&=&\!\!\!\!\int_{0}^{a}dz\left[{\lambda\over 8}\,G^{2}(x,x)
+{\delta m^{2}\over 2}\,G(x,x)+\delta\Lambda\right],
\nonumber \\
& &
\end {eqnarray}
where $G(x,x')$ is the Green function of the non-interacting theory, 
but obeying the BC, $\delta m^2$ is the radiatively induced shift in the 
mass parameter, and $\delta\Lambda$ is the shift in the cosmological constant 
(i.e., the change in the vacuum energy which is due solely to the
interaction, and not to the confinement). 

The spectral representation of the Euclidean Green function in 
$(d+1)$-dimensions is given by
\begin {eqnarray}
G(x,x')\!\!\!\!&=&\!\!\!\!\int {d\omega\over 2\pi}\int {d^{d-1}k\over(2\pi)^{d-1}}
e^{-i\omega(\tau-\tau')+i{\bf k}\cdot({\bf r}-{\bf r}')}
\nonumber \\
\!\!\!\!& &\!\!\!\!\times\sum_{n}{\varphi_{n}(z)\,\varphi_{n}^{*}(z')\over
\omega^{2}+k^{2}+m^{2}+k_{n}^{2}},
\end {eqnarray}
where, in the DN case, 
\begin {equation}
\label{varphi}
\varphi_{n}(z)=\sqrt{2\over a}\,\sin(k_nz),
\end{equation}
\begin{equation}
\label{k_n}
k_{n}=\left(n+\frac{1}{2}\right)\frac{\pi}{a}
\qquad(n=0,1,2,\ldots).
\end {equation}

Note that $G(x,x')$ diverges when $x'\to x$ for 
$d\ge 1$. Therefore,  a regularization
prescription is needed. Using dimensional regularization
we obtain
\begin {equation}
\label {zxc1}
G(x,x)={\Gamma\left(1-d/2\right)\over(4\pi)^{d/2}}
\sum_{n=0}^{\infty}\omega_{n}^{d-2}\,\varphi_{n}(z)\,\varphi_{n}^{*}(z),
\end {equation}
where $\omega_{n}=\sqrt{m^{2}+k_{n}^{2}}$.

Now we need to compute the terms appearing in Eq.\ (\ref{Einicial}).
Using the explicit form of $\varphi_n(z)$ given in Eqs.\ (\ref{varphi})
and (\ref{k_n}) one obtains
\begin{equation}
\int_{0}^{a}dz\,G(x,x)=
{\Gamma\left(1-d/2\right)\over(4\pi)^{d/2}}
\sum_{n=0}^{\infty}\omega_{n}^{d-2},
\end{equation}
\begin{eqnarray}
\int_{0}^{a}dz\, G^{2}(x,x)\!\!\!\!&=&\!\!\!\!
{\Gamma^{2}\left(1-d/2\right)\over (4\pi)^d\, a}
\bigg[\bigg(\sum_{n=0}^{\infty}\omega_{n}^{d-2}\bigg)^{2}
\nonumber \\
\!\!\!\!& &\!\!\!\!+{1\over 2}\sum_{n=0}^{\infty}\omega_{n}^{2d-4}\bigg].
\end{eqnarray}
Collecting terms, we obtain
\begin{eqnarray}
E_{\rm DN}^{(1)}\!\!\!\!&=&\!\!\!\!{\lambda\over 8a}\bigg[{\Gamma\left(1-d/2\right)\over(4\pi)^{d/2}}
\left[F(2-d,2a)\right.
\nonumber \\
\!\!\!\!& &\!\!\!\!\left.-F(2-d,a)\right]+{2a\,\delta m^{2}\over\lambda}\bigg]^{2}
\nonumber \\
\!\!\!\!& &\!\!\!\!+{\lambda\over 16a}{\Gamma^{2}\left(1-d/2\right)\over(4\pi)^{d}}
\left[F(4-2d,2a)\right.
\nonumber \\
\!\!\!\!& &\!\!\!\!\left.-F(4-2d,a)\right]
+\left[{\delta\Lambda}-{(\delta m^{2})^{2}\over 2\lambda}\right]a,
\label{E4passo}
\end {eqnarray}
where $F(s,a)$ is defined as
\begin {equation}
F(s,a):=\sum_{n=1}^{\infty}\left[m^{2}+\left({n\pi\over a}\right)^{2}\right]^{-s/2},
\end {equation}
for $\Re(s)>1$, and
can be extended analytically to the whole 
complex $s$-plane via the identity \cite{AmbjornWolfram83}
\begin {eqnarray}
\label{extF}
F(s,a)\!\!\!\!&=&\!\!\!\!-{1\over 2}m^{-s}
+{am^{1-s}\over 2\pi^{1/2}\,\Gamma\left(s/2\right)}
\bigg[\Gamma\left({s-1\over 2}\right)
\nonumber \\
\!\!\!\!& &\!\!\!\!+4\sum_{n=1}^{\infty}
{K_{(1-s)/2}(2nma)\over(nma)^{(1-s)/2}}\bigg].
\end {eqnarray}

In the above equation $K_{\nu}$ denotes the modified Bessel function 
of second kind. The structure 
of poles of $F(s,a)$ is dictated by the $\Gamma$ function: there are  
simple poles at $s=1,-1,-3,-5,\ldots$.

Let us now choose the renormalization conditions for $\delta m^2$ and 
$\delta\Lambda$. With this goal, recall that up to first order in 
$\lambda$ the self-energy is given by 
$\Sigma(x)=(\lambda/2)G(x,x)+\delta m^{2}$.  We shall fix $\delta m^{2}$ 
by imposing the following conditions 
on $\Sigma(x)$: (i) $\Sigma(x)<\infty$ (except possibly
at some special points); (ii) $\Sigma(x)$ vanishes
away from the plates when $a\to\infty$:
$\lim_{a\to\infty}\Sigma(z=\gamma a)=0$ for $0<\gamma<1$, and
(iii) $\delta m^2$ must  be independent of $a$.
These conditions are fulfilled by taking $\delta m^2=-(\lambda/2)\,G_0(0)$, 
where $G_{0}(z)$ denotes the non-interacting Green function 
{\em without boundary conditions} evaluated at the point $x=(0,{\bf 0},z)$. 

Computing $G_0(0)$ within dimensional regularization, we get
\begin {equation}
\label{dm2}
\delta m^2=-{\lambda\over 2}{\Gamma\left((1-d)/2\right)\over (4\pi)^{(d+1)/2}}\,m^{d-1}.
\end {equation}

\begin {figure}[!h]
\includegraphics*[scale=0.45, viewport=-5 0 1000 340]{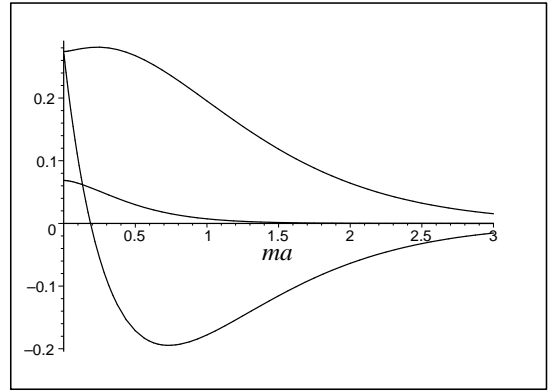}
\caption{$(512\pi^2/\lambda)\,a^3E^{(1)}$ (in $d=3$) as~a function of $ma$ for 
three kinds of
boundary conditions: Dirichlet-Dirichlet (upper curve), Neumann-Neumann 
(the curve that crosses the horizontal axis), and Dirichlet-Neumann.}
\label{3graficos}
\end {figure}

For the shift in the cosmological constant we shall take 
$\delta\Lambda=(\delta m^{2})^{2}/2\lambda$.
With this choice one eliminates the term proportional to $a$ in 
Eq.\ (\ref{E4passo}), which does not contribute
to the force between the plates (the linear dependence on $a$ is canceled
by similar terms when one adds the energy of the regions $z<0$
and $z>a$). Collecting all these results, we obtain
\begin {eqnarray}
E_{\rm DN}^{(1)}\!\!\!\!&=&\!\!\!\!{2\lambda a m^{2d-2}\over (4\pi)^{d+1}}
\bigg[\sum_{n=1}^{\infty}\bigg({2K_{(d-1)/2}(4nma)\over(2nma)^{(d-1)/2}}
\nonumber \\
\!\!\!\!& &\!\!\!\!-{K_{(d-1)/2}(2nma)\over(nma)^{(d-1)/2}}\bigg)\bigg]^{2}
\nonumber \\
\!\!\!\!& &\!\!\!\!+{\lambda\over 16}{\Gamma^{2}(1-d/2)\,m^{d-3}\over(4\pi)^{d+1/2}\,
\Gamma(2-d)}\bigg[\Gamma\left(\frac{3}{2}-d\right)
\nonumber \\
\!\!\!\!& &\!\!\!\!+4\sum_{n=1}^{\infty}\bigg({2K_{(2d-3)/2}(4nma)\over(2nma)^{(2d-3)/2}}
\nonumber \\
\!\!\!\!& &\!\!\!\!-{K_{(2d-3)/2}(2nma)\over(nma)^{(2d-3)/2}}\bigg)\bigg].
\end {eqnarray}
Taking $d=3$, we get
\begin {eqnarray}
\label {E1d=3DD}
E_{\rm DN}^{(1)}\!\!\!\!&=&\!\!\!\!{\lambda m^{2}\over 128 \pi^{4}a}
\bigg(\sum_{n=1}^{\infty}{K_{1}(4nma)-K_{1}(2nma)\over n}\bigg)^{2}.
\nonumber \\
& &
\end {eqnarray}
The results for the DD and NN BC (in $d=3$)
are given by \cite{FRF}
\begin {equation}
\label {defFDD}
E_{\rm DD}^{(1)}=\frac{\lambda m^2}{512\pi^2a}\bigg[\bigg(1+{2\over\pi}
\sum_{n=1}^{\infty}{K_{1}(2nma)\over n}\bigg)^{2}-1\bigg],
\end{equation}
\begin{equation}
\label {defFNN}
E_{\rm NN}^{(1)}=\frac{\lambda m^2}{512\pi^2a}\bigg[\bigg(1-{2\over\pi}
\sum_{n=1}^{\infty}{K_{1}(2nma)\over n}\bigg)^{2}-1\bigg].
\end{equation}
These results are graphically illustrated in
Fig.\ \ref{3graficos}.

It follows from Eqs.\ (\ref{defFDD}) and (\ref{defFNN}) that 
$E_{\rm DD}^{(1)}$ and $E_{\rm NN}^{(1)}$ are not equal, 
except in the zero mass case.
Indeed, using the expansion \cite{Braden82}
\begin{equation}
\sum_{n=1}^{\infty}{K_{1}(nz)\over n}={\pi^{2}\over 6z}
-{\pi\over 2}+O(z\,\ln z),
\end{equation}
valid for small $z$, we can take the limit $m\to 0$
in Eqs.\ (\ref{E1d=3DD})--(\ref{defFNN}), obtaining
\begin{equation}
E_{\rm DN}^{(1)}={\lambda\over 2^{13}3^{2}a^{3}},
\quad E_{\rm DD}^{(1)}=E_{\rm NN}^{(1)}={\lambda\over 2^{11}3^{2}a^{3}},
\label {EDD}
\end{equation}
which agree with the results obtained by Krech and
Dietrich \cite{Scalar} for $d=3$.\footnote{Note that
their $d$ equals our $d$ plus one.}

In the large mass limit ($ma\gg 1$) we have
\begin{equation}
\label{EDN1m}
E_{\rm DN}^{(1)}\approx\frac{\lambda m}{512\pi^{3}a^{2}}\,\exp(-4ma),
\end{equation}
\begin{equation}
\label{EDD1m}
E_{\rm DD}^{(1)}\approx-E_{\rm NN}^{(1)}
\approx\frac{\lambda}{256\pi}\left(\frac{m}{\pi a}\right)^{3/2}\exp(-2ma).
\end{equation}
Two aspects of the results above are worth of mention:
(i) while for $m=0$ $E_{\rm NN}^{(1)}$ is positive for all $a$, for
$m\ne 0$ it eventually becomes negative for sufficiently large $a$
(more precisely, for $a\approx 0.2 m^{-1}$);
(ii) while $E^{(1)}$ decays with distance (or mass) as fast as $E^{(0)}$
for DD or NN boundary conditions [cf.\ Eqs.\ (\ref{EDD0m}) and (\ref{EDD1m})],
the former decays faster than the latter in the DN case
[cf.\ Eqs.\ (\ref{EDN0m}) and (\ref{EDN1m})].

\section{FINAL REMARKS}

We have computed the first radiative correction to the Casimir energy of 
the massive $\lambda\phi^4$ model subject to three distinct BC on a pair
of parallel plates. We showed  
that while for a massless field DD and NN boundary conditions lead to the same 
$O(\lambda)$ radiative correction (an unexpected result), this 
is not true for a massive field. In addition, that correction presents very
distinct behavior as a function of the distance $a$ between the plates:
while $a^3E_{\rm DD}^{(1)}$ first increases and then decreases with $a$,
$a^3E_{\rm NN}^{(1)}$ first decreases and then increases with $a$ 
(see Fig.\ \ref{3graficos}). As a consequence, a pair of DD plates
is more attracted to each other than a pair of NN plates
when the distance between the plates is sufficiently small;
the opposite occurs when the plates are far apart.

We also computed for the 
first time the $O(\lambda)$ radiative correction to the Casimir
energy for that model subject to DN boundary conditions 
(i.e., Dirichlet BC on one plate 
and Neumann BC on the other). Our results show that in this case
the correction to the one-loop result
is much smaller than in the DD or NN cases for large separations
between the plates (i.e., for $a\gg m^{-1}$).

Results for other kinds of BC (periodic and anti-periodic) 
will be presented elsewhere. 

\section*{ACKNOWLEDGMENTS}

This work was supported by CNPq and CAPES.

\end{document}